\documentclass{appolb}
\usepackage{braket}
\usepackage{epsfig}
\usepackage{graphicx}
\usepackage[english]{babel}
\usepackage{hyphenat}
\usepackage{amsmath}
\usepackage{amssymb}
\usepackage{mathtools}
\usepackage{mathrsfs}
\usepackage{slashed}
\usepackage{wrapfig}
\usepackage{epstopdf}
\usepackage{xcolor}
\usepackage{booktabs}
\usepackage{subfigure}
\graphicspath{ {image/} }
\usepackage{float}
\usepackage{ulem}
\definecolor{lcolor}{rgb}{0.,0.0,0.}
\definecolor{citcolor}{rgb}{0,0.,0.5}
\usepackage[breaklinks,colorlinks,urlcolor=blue,citecolor=blue,linkcolor=blue]{hyperref}
\usepackage{multirow}
\usepackage{ltablex}
\usepackage{soul}
\usepackage{hepunits}

\def\rrangle{\rangle\!\rangle}
\def\llangle{\langle\!\langle}
\newcommand{\beq}{\begin{eqnarray}}
\newcommand{\eeq}{\end{eqnarray}}

\newcommand{\bem}{\begin{multline}}
\newcommand{\eem}{\end{multline}}
\newcommand{\beg}{\begin{gather}}
\newcommand{\eeg}{\end{gather}}

\newcommand{\ben}{\begin{eqnarray*}}
\newcommand{\een}{\end{eqnarray*}}

\newcommand{\secn}[1]{Section~1}
\newcommand{\appn}[1]{Appendix~1}

\long\def\comment#1{ }

\def\and{\quad\text{and}\quad}

\def\0{{\boldsymbol 0}}
\def\p{{\boldsymbol p}}

\def\x{{\boldsymbol x}}
\def\y{{\boldsymbol y}}

\begin{document}
\title{A first step towards quantum simulating jet evolution in a dense medium%
\thanks{Presented at Quark Matter 2022}%
}
\author{João Barata, 
\address{Physics Department, Brookhaven National Laboratory, Upton, NY 11973, USA}
}
\maketitle
\begin{abstract}
    The fast development of quantum technologies over the last decades has offered a glimpse to a future where the quantum properties of multi-particle systems might be more fully understood. In particular, quantum computing might prove crucial to explore many aspects of high energy physics unaccessible to classical methods. In this talk, we will describe how one can use digital quantum computers to study the evolution of QCD jets in quark gluons plasmas. We construct a quantum circuit to study single particle evolution in a dense QCD medium. Focusing on the jet quenching parameter $\hat q $, we present some early numerical results for a small quantum circuit. Future extensions of this strategy are also addressed.
\end{abstract}
  
\section{Introduction}
Heavy ion collisions at RHIC and the LHC have offered an unique opportunity to explore the quark gluon plasma, leading the way towards a better understanding of QCD. Experimentally, there are multiple types of probes used to indirectly measure the properties of this state of matter, being jets one of the most powerful ones. When  traversing the plasma, jets' constituents suffer multiple interactions with the background plasma, leading to their trajectories being modified and resulting in the production of extra soft radiation~\cite{Blaizot:2015lma}. 

Understanding how such modifications come into place allows to infer the properties of the underlying plasma. However, from a theoretical point of view it is hard to study these type of effects once the particle number becomes large. Conversely, many of the interesting medium induced effects only emerge at higher particle multiplicities, and thus understanding how multiparticle interference mechanisms come into play is of great importance. 

Quantum computers are quantum machines which ideally can simulate the full quantum dynamics of other target quantum systems. Thus, in principle, one could use such devices to simulate jet evolution in the presence of a background, capturing higher order interference effects. In this way, quantum computers might allow to broaden our understanding of jet quenching physics. However, in reality, current quantum computers are still relatively small and imperfect machines, and thus they are not capable of tackling such complex questions~\cite{Preskill_2018}. Moreover, it is still not even clear how to efficiently (i.e. using less resources than a classical machine) formulate full jet evolution at the level of quantum operations on the computer.

In this work we detail our current effort to implement jet in-medium evolution in a dense stochastic background. At this point, we limit our discussion to the evolution of a single quark state, neglecting radiation production. This first step will be useful when including gluons, since many aspects of the implementation are common. 

\section{Jet evolution in a dense medium and the quantum simulation algorithm}
We consider the evolution of a single quark (the jet) near the light cone with momentum $p^\mu=(p^+,p^-,\p)$. The medium is described by a classical background field $A_\mu^a$. In the usual picture, the jet is highly energetic and thus, in the light-cone gauge $A^+=0$, it is only sensitive to the $A^-$ component of the field and its spacetime dependence can be simplified as $A(x^\mu)\approx A(x^+,\x) $. One can then show that the quark evolves according to an effective Hamiltonian~\cite{Blaizot:2015lma,Li:2020uhl}
\begin{align}
    \hat H = \hat K+ \hat V =  \frac{\hat \p^2}{2p^+} +g \hat A^-_a T^a\; ,
\end{align} 
at constant $p^+$ (the jet energy). The first term corresponds to the kinetic energy of the quark and it accounts for the quantum diffusion of the probe in the medium. The potential terms takes into account the gluon exchanges between the medium and the quark. 

In this approach, the medium is described by a stochastic field. As a consequence, when computing any observable, one must average over all possible medium configurations. In practice, we simulate the evolution of the system over an ensemble of field configurations, such that for each configuration the final quantum state is given by 
\begin{align}\label{eq:master_ev}
    \begin{split}
     \ket{\psi_{L}}=& U(L;0)\ket{\psi_0} \;,
    \end{split}
\end{align}
with $U$ the time evolution operator. Each field is generated by solving the classical Yang-Mills equations in the presence of a stochastic color source $\rho^a$, which is assumed to satisfy a Gaussian error form, i.e. its only non-trivial correlator is
\begin{align}\label{eq:rhorho}
    \llangle \rho_a(x^+,\x)\rho_b(y^+,\y) \rrangle =g^2 \mu^2(\x)\delta_{ab}\delta^2(\x-\y)\delta(x^+-y^+)\; ,
\end{align}
where $g^2 \mu^2$ should be understood as the medium density. 

In this work we are interested in computing the squared transverse momentum acquired by the quark due to evolution in the medium.  The medium averaged expectation value of this observable can be related to the jet quenching parameter $\hat q$ via
\begin{align}\label{eq:def_qhat}
    \hat q \equiv \frac{\llangle \langle \hat \p^2 (L)  \rangle \rrangle-\llangle \langle \hat \p^2 (0)  \rangle \rrangle }{L}  \, ,
    \end{align}
where $L$ is the longitudinal length of the medium. Here we denote the medium average as
\begin{align}\label{eq:medium_avg}
    \llangle \langle {\hat \p^2} \rangle  \rrangle =\lim_{n\to \infty}\frac{1}{n}\sum_{i=1}^n\langle {\hat \p^2} \rangle _i\; ,
\end{align}
where the sum runs over different mediums configurations generated according to Eq.~\ref{eq:rhorho} and the single brackets denote quantum expectation values.

To implement the evolution of the jet in a quantum computer we use the quantum simulation algorithm~\cite{nielsen_chuang_2010}. Its main features can be summarized as follows:
\begin{enumerate}
    \item \textbf{Input}: A description of the target quantum system in terms of the system Hamiltonian $H$ and underlying Hilbert space.

    \item \textbf{Encoding}:  Mapping the degrees of freedom of the problem to qubits on the quantum computer. One needs a mapping between the physical operators and quantum gates that act on the quantum circuit. 
    
    \item \textbf{Initial state preparation}: Preparing the initial state of the system from a fiducial state given by the computer. In the current study, these states are identified.

    \item \textbf{Time evolution}: After having prepared the initial state, one applies the time evolution operator on the computational state, in terms of a series of quantum gate operations.

    \item \textbf{Measurement}: Measurement of the final state, usually in such a way that relevant correlators can be extracted efficiently.
\end{enumerate}

\section{Quantum circuit and results}
\begin{wrapfigure}{L}{.5\textwidth}
        \centering
        \includegraphics[width=5.cm]{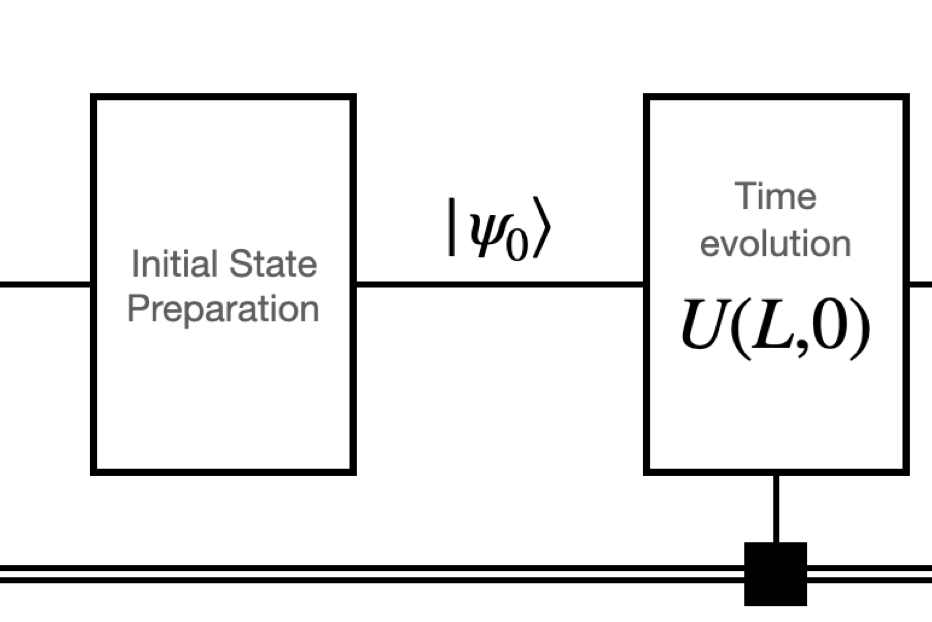}
    \caption{Circuit implementing the evolution of the quark in the medium. More details can be found in Ref.~\cite{Barata:2021yri}.}
        \label{Fig:circ}
    \end{wrapfigure}

The quantum circuit implementing the evolution of the quark according to the quantum simulation algorithm is depicted in Fig.~\ref{Fig:circ}. There, the double lines indicate a classical chanel of information which feeds into the quantum gate implementing the evolution operator. Through this chanel, for each simulation, one feeds the classical  field values to the proper quantum circuit. 

The time evolution operator is decomposed into multiple small time evolution operators. For each one of these small step evolutions, we further decompose the evolution into a term controlled by the kinetic operator and the potential term; for more details see Refs.~\cite{future, Barata:2021yri}. Finally, since we are working with a digital computer, one must provide a finite description of the problem. For that, we introduce a transverse lattice with position space spacing $a_\perp$. Using a binary mapping, we can then match the momentum/spatial modes of the probe to the qubits in the computer~\cite{future}.

From the simulations, we extract the underlying transverse momentum distribution of the quark by directly measuring the full quantum state. Then from the constructed distribution one can compute $\hat q$ according to Eq.~\ref{eq:def_qhat}. Such an approach is inefficient and more optimized forms of extracting correlators can be devised, see e.g. Refs.~\cite{future, Barata:2021yri} for further discussion.

\begin{figure}[htb]
    \centerline{%
    \includegraphics[width=5.cm]{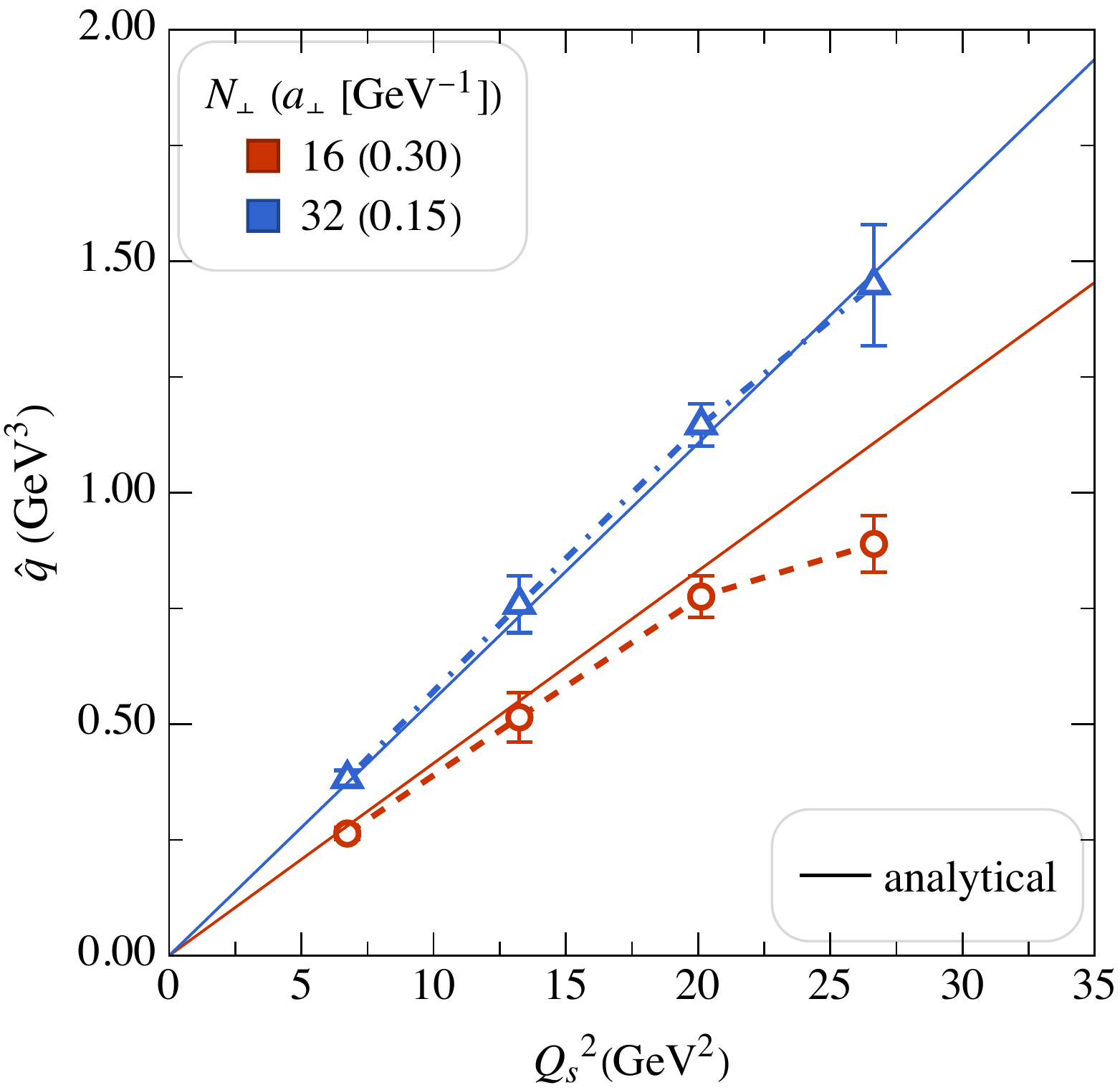}
    \includegraphics[width=5.cm]{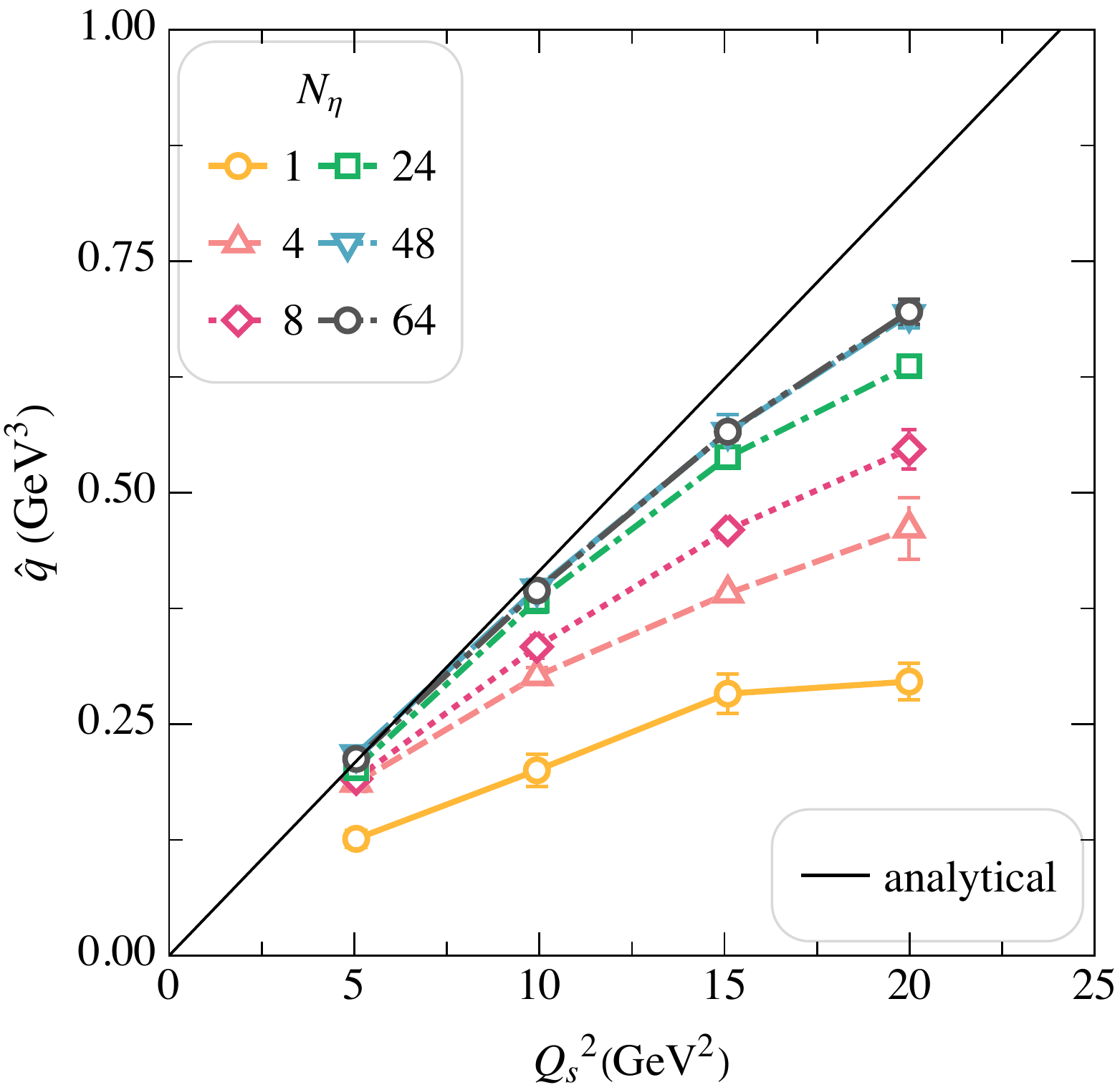}}
    \caption{Left: Jet quenching parameter as a function of the saturation scale for two different lattices in a U(1) medium at infinite jet energy. Right: The same observable for a SU(2) medium, using the coarser lattice and implementing different values for $N_\eta$ at finite jet energy. Figures taken from Ref.~\cite{future}.}
    \label{Fig:qhat_1}
\end{figure}

In Fig.~\ref{Fig:qhat_1} (left), we show the results obtained for the evolution of the jet quenching coefficient $\hat q$ as a function of the saturation scale for a U(1) background with $p^+=\infty$. We perform the simulation using two transverse lattices: one with lattice spacing in position space $a_\perp=0.30 \, {\rm GeV}^{-1}$ and a finer one with $a_\perp=0.15 \, {\rm GeV}^{-1}$. In practice, the difference between these two lattices is related to their sensitivity to discretization effects, with the finer lattice being less sensitive. Indeed, we observe this in Fig.~\ref{Fig:qhat_1} (left), with the coarser lattice showing a non-linear evolution at large values of $Q_s^2=C_F g^4 \mu^2 L/(2\pi) $, which are not expected when compared to the analytical calculation of $\hat q $~\cite{future,Blaizot:2015lma}. When moving to the finer lattice, we see that indeed these effects disappear and one recovers the expected behavior.

\begin{wrapfigure}{L}{.5\textwidth}
    \centering
    \includegraphics[width=5.cm]{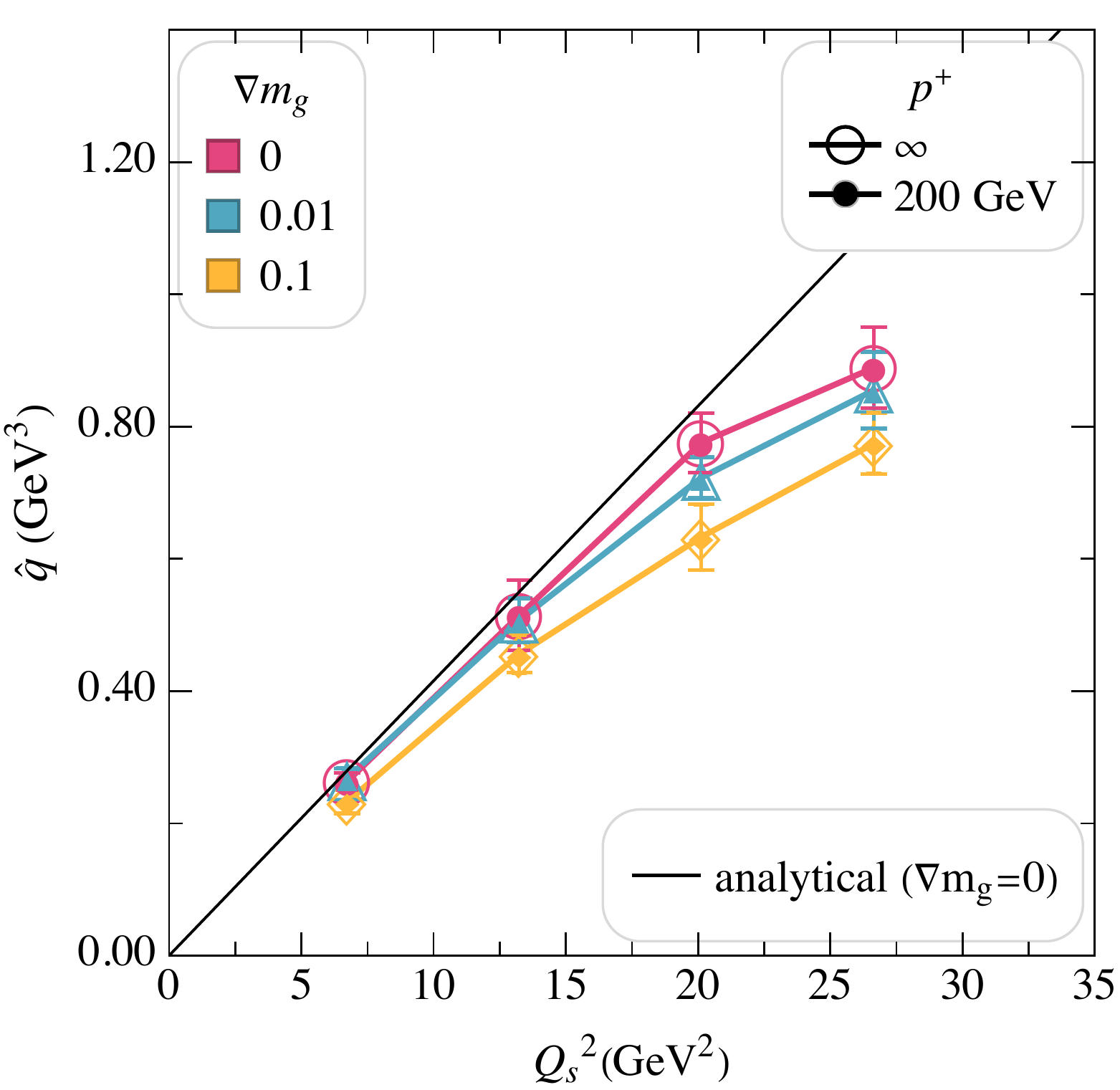}
    \caption{Jet quenching parameter as a function of the saturation scale for an anisotropic medium. Figure from Ref.~\cite{future}.}
    \label{Fig:qhat_an}
\end{wrapfigure}

The same exercise can be performed for a SU(2) medium at finite jet energy ($p^+=200$ GeV). The results are shown in Fig.~\ref{Fig:qhat_1} (right) for the same coarse lattice. For these simulations, one needs to do an extra subdivision of the evolution operator in order to ensure that Eq.~\ref{eq:rhorho} is satisfied. For that we introduce $N_\eta$
layers, within each we generate a novel field configuration. Indeed, we observe in the numerical result that as the number of layers increases, the the curves tend to converge towards the analytical result (modulus lattice effects).

Finally, the current approach can be used beyond the simplest models of the medium. An example of an interesting scenario, which has been recently explored~\cite{Barata:2022krd,Fu:2022idl,Sadofyev:2021ohn}, is to consider a medium with anisotropic transverse profile. In Fig.~\ref{Fig:qhat_an} we show the results for $\hat q$ in the presence of a medium where the Debye mass $m_g(\x)=m_g(1+\nabla m_g\cdot \x)$. Then, varying the value of $|\nabla m_g|$, we observe that $\hat q$ gets corrected with respect to the homogenous case. In particular, these modifications are not energy suppressed, unlike the ones considered in Refs.~\cite{Barata:2022krd,Fu:2022idl}. 
In conclusion, our approach offers a way to explore this type of novel effects and gauge their importance for jet quenching phenomenology.

\section{Conclusion}
In this proceeding, we have shown that the evolution of a single particle in a QCD background can be studied using a quantum computer. Our results reproduce known analytical results, but also point towards applications of this approach to more realistic scenarios hard to access analytically. In the future, we plan to extend this program to include the production of soft gluon radiation.

\section{Acknowledgements}
We are grateful to X. Du, M. Li, C. A. Salgado and W. Qian, who have made important contributions to Ref.~\cite{future}. J.B.’s work is supported by the U.S. Department of Energy, Office of Science, National Quantum Information Science Research Centers under the “Co-design Center for Quantum Advantage” award and by the U.S. Department of Energy, Office of Science, Office of Nuclear Physics, under contract No. DE-SC0012704.

\bibliographystyle{plain}
\bibliography{Lib.bib}

\end{document}